\documentclass{article}
\usepackage{graphicx}
\begin{document}

\makeatletter	   
\renewcommand{\ps@plain}{%
     \renewcommand{\@evenhead}{\@oddhead}%
     \renewcommand{\@oddfoot}{}
     \renewcommand{\@evenfoot}{\@oddfoot}}
\makeatother     
\title{Gravity from Discrete Interactions,\\
a laboratory of ideas for field theory}
\author
{Manoelito M de Souza\\
              Universidade Estadual do Sudoeste da Bahia \footnote{Permanent address}\\
Departamento de Ci\^encias Exatas - Vit\'oria da Conquista - BA - Brazil\\
             and \\ 
Universidade Federal do Esp\'{\i}rito Santo\footnote{actual address}\footnote{manoelitosouza@yahoo.com}\\ 
Departamento de F\'{\i}sica - Vit\'oria - ES - Brazil\\}

\date{}          
\maketitle

\pagestyle{plain}

\begin{abstract}
The observed accelerated expansion of the universe is an indication that somehow, in some conditions, gravity may become a repulsive interaction. The very concept of discrete interactions, compatible with General Relativity as an effective theory of averaged fields, can handle this and other pressing problems without the need of searching for ad hoc exotic sources. Semi-quantum models of discrete interactions based on quantum exchanges between point-like masses are a rich laboratory for enlightening field interactions and classical-to-quantum transitions in field  theory.  A very simple  version is presented for exhibiting how  naturally it makes contact with inflation, dark matter and dark energy issues.

\end{abstract}
\section{Introduction}
In  few more years the scientific community will be celebrating the centennial year of the two great pillars of modern theoretical physics, Quantum Mechanics and General Relativity, without being able to make them compatible, harmonize them into a single scheme. Quantum Mechanics carries the idea of interaction discreteness realized through  exchanges of a quantum interaction, a packet of energy and momentum, while General Relativity associates the gravitational interaction to the smooth geometry of a curved spacetime. This incompatibility is a serious problem because they are both unbeatable on fitting the experimental data with high accuracy on their respective domain of validity and applicability. General Relativity overseeds the old Newtonian scheme  reproducing its dynamics in its Newtonian limit, a non-relativistic weak field limit. There are great challenges facing General Relativity \cite{Perlmutter}. 
The observed accelerated expansion of the universe is an indication that somehow, in some conditions, gravity may become a repulsive interaction. The old question of the flat rotation curves of stars and galaxies is still  unsettled. These questions have in common  very large scales of distances which lead to the above mentioned weak field Newtonian limit.\footnote{The flat rotation curves occurs when the field is smaller than $10^{-9}m/s^2$ which, by the way, is of the same order of the Pioneer anomaly \cite{gr-qc/0104064}. This may not be a mere coincidence.}  In the Newtonian scheme there were no room for a repulsive gravity but General Relativity, assuming its validity on such scales of distance, can handle it in various ways but none without  controversy.  The Einstein field equations are the equality between, in one side, the geometric one representing the observational data, the field, the test mass acceleration, and, in the other side, the sources, the matter content represented by the stress-energy tensor. Feeding this scheme with the observed data for finding the matter content, an exotic fluid that fulfill this data, may be  an easy but certainly dangerous procedure as it raises the Popper's  falsifiability question if there is no pre-established  criteria on the acceptability of what is found. 
The same observations are valid for the galactic flat rotation curves. Then we have the dark matter and the dark energy controversies.
Inflation in cosmology, its nature, what powers it and  what smoothly disconnects it is another controversy, or unsettled question in gravitation theory. These are certainly among the greatest riddles for the modern researcher in gravitation theory.

 The objective of this letter is to point to a new  remarkable fact that they can altogether be enlightened with just very simple ideas on discrete interactions and no further else assumption. For discrete interaction it is meant the instantaneous emission/absorption of a quantum of interaction triggered by  the instantaneous absorption/emission of a quantum of interaction, with free, straight-line propagation  between any two consecutive interactions, in a flat spacetime background. As an illustration we present here the  simplest discrete interaction model for gravity between two point-like masses in a non-relativistic relative radial motion; it can easily be associated to a cosmological model of a spherically symmetric distribution of dust. 
 Such kind of naive models, with no free parameters and great variety of levels of complexity, are  useful laboratories of ideas in the search for a better understanding of field interactions and quantum to classical transitions. The resulting  dynamics,  strongly dependent on the distance scales and on the initial conditions,  oscillates between repulsive and attractive phases. It starts with a repulsive exponential expansion that smoothly changes into a Newtonian phase to return later, again to a repulsive expansion.  This rich scenario needs no extra input to develop besides the two point-like masses and discrete interactions.  The well known success of General Relativity on high accurately fitting some experimental data does not stand against gravity being a discrete interaction since the very General Relativity \cite{gr-qc/9801040}, like Classical Electrodynamics, can be seen as an effective theory of averaged interactions in a discrete interaction context  \cite{hep-th/9610028,BJP,BJP2}. They are both finite  discrete interaction theories;  singularities and other well known problems are shown to be consequences of the averaging process in the field definition.

\section{The model}
Let us consider two point-like masses, M and m,  in relative 
non-relativistic radial motion under a discrete gravitational interaction. $m$ is a test mass ($M>>m$) and its momentum and position (relative to $M$) are given by
\begin{equation}\label{pnd}
p_n=p_{n-1}+\Delta p_{n-1}=p_0+\sum_{j=0}^{n-1}\Delta p_{j}, \end{equation}
as the momentum only changes at discrete points when there is the emission/absorption of an interaction quantum, a graviton, and
\begin{equation}\label{rn}
r_{n}=r_{n-1}+\Delta r_{n-1}=r_{n-1}+\frac{p_{n-1}}{m }\Delta t_{n-1}=r_0+\sum_{j=0}^{n-1}\frac{p_j\Delta t_j}{m},
\end{equation}
as there is free  straight-line propagation between any two consecutive interactions.
The label $n$ counts the number of discrete interactions starting from $t_0$ at the initial position $r_0$ with the initial momentum $p_0.$ With such discrete interactions a trajectory is  piecewise-linear with vertices at the interaction events, where  velocity, as the trajectory tangent vector, is not defined because the discrete interaction causes a sudden change of velocity. So the notion of acceleration, if not as an approximation, is meaningless since it is null in the free propagation period between two consecutive discrete interaction and is not defined at the vertices. 
\subsection*{The kinematics}

Among many other alternatives,  we take for our model of classical discrete interaction
\begin{equation}
\label{dtn}
\Delta t_n=\alpha x_{n}
\end{equation}
where $x_n=\frac{r_n}{r_0}$ and $\alpha$ is the kinematic coefficient that describes the system free propagation between any two consecutive interactions. This choice is justified  assuming that, let us say, the graviton, propagates with the speed of light $c$ and, crudely, that the emission of an interaction quantum is triggered without delay by the absorption of an interaction quantum. Then, in the non-relativistic approximation
\begin{equation}\label{alpha}\Delta t_n\approx \frac{2r_{n}}{c}=\frac{2r_{0}}{c}x_n,\qquad\alpha\approx \frac{2r_{0}}{c}.\end{equation}

\subsection*{The dynamics}
The dynamics for our model will be chosen with an eye on the above mentioned non-relativistic weak field limit, which, after (\ref{dtn}), implies on
\begin{equation}
\label{dpn}\Delta p_n=\beta/ x_{n},
\end{equation}
as the most immediate choice, since 
\begin{equation}
\label{deltaed}\frac{\Delta p_n}{\Delta t_n}\approx\frac{d p_n}{d t_n}\end{equation}  requires
\begin{equation}
\frac{\beta}{\alpha}\frac{1}{x_n^2}\equiv-\frac{GMm}{r_n^2}=-\frac{GMm}{r_0^2}\frac{1}{x_n^2},\end{equation}
which fixes the dynamic coefficient $\beta$ as

\begin{equation}
\beta=-\frac{2GMm}{c r_0}.\end{equation} There is no free-to-be-adjusted constant in this toy model of discrete classical interactions between two point-like masses.
However, we strongly remark that (\ref{dpn}) is not the unique possible choice, even after (\ref{dtn}), and neither the most appropriate considering that the Newton's equation is just an approximation to a correct formulation of the problem. It is, nonetheless, interesting to see that discrete interaction  can produce repulsive gravity even in such a quasi-Newtonian simple context.

On the other hand, (\ref{deltaed}) has a price as it requires an approximate continuity on $t_n$ and $p_n$ (grossly   $\Delta t_n<<1$ and $\Delta p_n<<1$) which, with (\ref{dtn}) and (\ref{dpn}), leads to 
\begin{equation}
\label{range}
\beta<<x_n<<\frac{1}{\alpha},
\end{equation}
that, for dimensional reasons, must be seen just as a qualitative indication of  the existence of such a range of validity for the Newton`s law of gravity.
\subsection*{A word of caution}
A word of caution, with respect to (\ref{alpha}), is however necessary. We are dealing here with the gravitational interaction between two point-like mass. In practice this would imply on something like the gravitational interaction between two electrons, or neutrinos or some other elementary particle. For a  long time still to come this would be a purely academic business! 
If $M$ is due to a number $N_M$ of point-like 
masses distributed on a region of radius $r_M$, with $r_M<<r_0$, and uncorrelatedly interacting with $m$, then 
$\alpha=\frac{2r_0}{c}$ must be replaced by
\begin{equation}
\label{dtnN}
\alpha=\frac{2r_0}{cN_M}.
\end{equation}
Just for an order of magnitude, we remind that light takes about 8 minutes to reach Earth from the Sun which cannot, obviously, be the order of the $\Delta t_n$. $N_M$ for the Sun is such a huge number that most certainly we will never be able to measure such a tiny lapse of time $\Delta t_n$.
On the other hand, in this non-relativistic approach $\alpha$ is the only place where $c$, the speed of light, will appear. We will keep $N_M=1$ everywhere and in the case we need to consider a real physical observation involving a composed object we just have to replace $c$ by $cN_M $:
\begin{equation}
\label{cNM}
c\Rightarrow cN_M 
\end{equation}
 In general we do not  know  $N_M$ but we can assume, in a gross approximation, that
\[N_M\propto M.\]
A similar reasoning \cite{else} must be considered at the moment that $m$ absorbs an interaction quantum and if $m$ is composed of $N_m$ point-like objects then the dynamic coefficient $\beta$ (or the probability that this absorption occurs) is  proportional to $N_m$ and then to $m$. This justifies  part of the Newton's statement  on the gravitational force (being proportional to the product of the masses).

\section{Solving for the equation of movement}

 After (\ref{dtn})) and (\ref{dpn}), with $a\equiv\frac{p_0}{m r_{0}}=\frac{v_0}{r_0}$ and $b\equiv \frac{\beta}{\alpha m r_{0}}=-\frac{GM}{r_0^3}=-\frac{v_{E,0}^2}{2r_0^2},$ (\ref{rn}) may be written as
\begin{equation}
\label{xn}
x_{n}=1+ \frac{\alpha}{mr_{0}}\sum_{j=0}^{n-1}(p_0 +\sum_{j_1=0}^{j-1}\frac{\beta}{x_{j_1}})x_{j}=1+ \sum_{j=0}^{n-1}( \alpha\, a+\sum_{j_1=0}^{j-1}\frac{b\,\alpha^2}{x_{j_1}})x_{j},\end{equation} 
which can be solved by re-iterations but it is more convenient to use
 an expansion  in series powers of $\alpha$ 
\begin{equation}
\label{xps}
x_n=\sum_{s\ge0}\alpha^s x_{n,s},\;\;\; x_{n,0}=1,
\end{equation}
where $x_{n,s}$  are  coefficients to be determined in terms of the initial conditions $r_0,$  $p_0$ and  of $n$. We find \cite{else}

$$x_{n,0}=1,\quad x_{n,1}=a{n\choose 1},\;\;\;\;\;\; x_{n,k}=0\;\;\;\textnormal{ for}\;\;\; k>[n/2]$$
and for $ k<[n/2]$:
\begin{equation}
\label{x2k}
x_{n,2k}=BA^{k-1}{n\choose 2k}+2(k-1)b BA^{k-2}{n\choose 2k-1}+\dots
\end{equation}
\begin{equation}
\label{x2kp1}
x_{n,2k+1}=aA^{k}{n\choose 2k+1}+a b A^{k-2}\left( k A+B(k-1)\right){n\choose 2k}+\dots,
\end{equation}
where the dots indicate  decreasing sequences of $n$-binomial numbers.
We have introduced, for convenience, the two constants $A$ and $B$, defined as
\begin{equation}
\label{A}
A\equiv a^2+2b=\frac{2}{m r_0^2}\left(\frac{p_0^2}{2m}-\frac{GMm}{r_0}\right)=\frac{v_0^2-v_{E,0}^2}{r_0^2}=\frac{w_0^2}{r_0^2},\end{equation}
 $$B=A-b=\frac{w_0^2+v_{0}^2}{2r_0^2},$$
\noindent where $v_0$ and $v_{E,0}$ are, respectively, the  velocity and the escape velocity at the initial position $r_0$ in the standard Newtonian potential;  $w_0^2\equiv v_0^2-v_{E,0}^2$.  $A$ is clearly related to the system initial energy. 

\section{The $n$-dominant order interaction.} 

If we  consider in each $x_{n,s}$ only the contributions from the term with the highest power in $n,$  ($n>>1$), 
 \begin{equation}
{n\choose {k}}=\frac{n^{k}}{(k)!}+0(n^{k-1})
\end{equation}
we have
  
\begin{equation}
x_{n,2k}\approx A^{k-1}B\frac{n^{2k}}{(2k)!}\qquad{for}\qquad k>0 \end{equation}

\begin{equation}
x_{n,2k+1}\approx aA^k\frac{n^{2k+1}}{(2k+1)!} \qquad{for}\qquad k\ge0,\end{equation}

and then,  from (\ref{xps}), assuming $A\not=0$,
\begin{equation}
\label{eq:Eq-13}
x_n\approx 1+\sum_{k\ge1}^{[n/2]}\alpha^{2k} A^{k-1}B\frac{n^{2k}}{(2k)!}+\sum_{k\ge0}^{[n/2]}\alpha^{2k +1}aA^k\frac{n^{2k+1}}{(2k+1)!}= 
\end{equation}

\[=1+\sum_{k\ge1}^{[n/2]}\frac{B}{A} \frac{X_{n}^{2k}}{(2k)!}+\sum_{k\ge0}^{[n/2]}\frac{a}{\sqrt A}\frac{X_{n}^{2k+1}}{(2k+1)!},\]
with $X_n=\alpha\sqrt{A}n=\frac{2w_0}{c}n$,
$n>>1,$ and $\alpha\sqrt{A}=\frac{2w_0}{c}<<1.$  It is a good approximation to replace the partial sums by their respective asymptotic  series, 

\begin{equation}
\label{xnh1}
x_n\approx\frac{b}{A}+\frac{B}{A}\cosh X_{n}+\frac{a}{\sqrt A}\sinh X_{n}
\end{equation}
 
\begin{equation}\label{xnh}
r_n\approx \frac{GM}{w_0^2}(\cosh\phi_n-1),
\end{equation}
with 
$ 
\phi_n\equiv X_{n}+\phi_0=\frac{2w_{0}n}{c}+\phi_0
$ and 
$$ 
\tanh\phi_0=\frac{a\sqrt A}{B}=\frac{2v_0 w_{0}}{w_{0}^2+v_0^2}.$$
 Replacing in (\ref{pnd})  the sum by an  integral, which is proved  to be valid \cite{else} at this order of approximation, leads  to
\begin{equation}\label{q}
q\equiv p_n-k_n=p_0-k_0, \quad \textnormal{with} \quad k_n=2 m w_0(e^{\phi_n}-1)^{-1}.
\end{equation}
$q$, which has the physical dimension of linear momentum, is a conserved quantity as it does not change with $n$. It would be kind of peculiar to have linear momentum  conservation under gravitational interaction but actually 
 it boils down \cite{else}  to  energy conservation 
\begin{equation}
q= m w=m\sqrt{v_n^2-\frac{2GM}{r_n}}=m\sqrt{v_0^2-\frac{2GM}{r_0}}=\sqrt{2mE}.
\end{equation}
The dominant-order contributions reproduce the usual Newtonian picture of continuous interactions with its known interaction potential. \\
 
\subsection*{The ``potential energy"}

The ``potential energy",  retrieved from the $q$ conservation 
\[U_n\equiv\frac{q^2-p_n^2}{2m}=\frac{(q-p_n)(q+p_n)}{2m}=\frac{k_n(2q+k_n)}{2m}\],
 
  can  be seen from a new perspective: we can be interested on knowing the zeroes of the potential, $2q+k_{\bar n}=0$ and $k_{\bar n}=0$.
 In this  dominant order contribution approximation each condition equally leads, with (\ref{q}),  to the same point at infinity
\[v_{E,\bar{n}}=0,\qquad\textnormal{or, equivalently}\qquad
1/r_{\bar n}=0,\]
as expected once the potential is the old Newtonian one, 
\[U_n=\frac{m^2(v_{n}^2-v_{E,n}^2)-p_n^2}{2m}=-\frac{mv_{E,n}^2}{2}= -\frac{GMm}{r_n}.\] 
But it is  remarkable the possibility of having  two distinct zeroes (this can happen with the inclusion of sub-dominant contributions) and then there is a place for a repulsive gravity too, as we will see next.
\section{Sub-dominant contributions. }

It is convenient now, for $w_0\not=0$, to combine the initial condition parameters $w_0$,  $v_0$ ($v_0>0$) and $r_0$ (or $v_{E,0}$) into a single one $h$, $h\equiv\frac{v_0}{|w_0|}$
\begin{equation}
\label{h}
h \equiv \cases{\frac{v_0}{\sqrt{v_0^2-v_{E,0}^2}}=\frac{1}{\sqrt{1-(\frac{v_{E,0}}{v_0})^2}}\ge1,&if $v_0>v_{E,0}$\cr
 \frac{v_0}{\sqrt{v_{E,0}^2-v_0^2}}=\frac{1}{\sqrt{(\frac{v_{E,0}}{v_0})^2-1}}\ge0,&if $v_0<v_{E,0}$\cr}
\end{equation} 
As $h\ge1$ is finite  with the equality to 1 being only asymptotically reached for 
\[v_{E,0}\rightarrow0\,\,\,\,or\,\,\,r_0\rightarrow\infty,\] 
we will be assuming from now on  $h>>1$ except when the opposite is explicitly stated. 

Including now the next $n$-order contributions i.e. the terms with the second highest power of $n$ in each $x_{n,s}$, 
\begin{equation}\label{sda}
{n\choose {k}}=\frac{n^{k}-{k\choose {2}}n^{k-1}}{(k)!}+0(n^{k-2})\end{equation}
we have, instead of  (\ref{x2k}) and (\ref{x2kp1}),
\[x_{n,2k}=BA^{k-1}\frac{n^{2k}-k(2k-1)n^{2k-1}}{(2k)!}+2(k-1)bBA^{k-2}\frac{n^{2k-1}}{(2k-1)!}+\dots\]
\[=\frac{B}{A}\left(\frac{(\sqrt{A}n)^{2k}}{(2k)!}-\frac{b}{\sqrt A}\frac{(\sqrt{A}n)^{2k-1}}{(2k-1)!}-\frac{na^2}{2}\frac{(\sqrt{A}n)^{2k-2}}{(2k-2)!}\right)+\dots\]
and  
  \[x_{n,2k+1}=aA^k\left(\frac{n^{2k+1}-k(2k+1)n^{2k}}{(2k+1)!}\right)+abA^{k-2}\left(kA+(k-1)B\right
)\frac{n^{2k}}{(2k)!}\]

\[=\frac{a}{\sqrt A}\left(\frac{(n\sqrt A)^{2k+1}}{(2k+1)!}-\frac{bB}{A^{\frac{3}{2}}}\frac{(n\sqrt A)^{2k}}{(2k)!}
+\frac{n  B^2 }{2A}\frac{(n\sqrt A)^{2k-1}}{(2k-1)!}\right)+\dots\]
and they lead, with (\ref{xps}),  to
\begin{equation}
\label{xna2}
x_n=T_0+T_1\cosh X_n+T_2\sinh X_n+\frac{1}{2}X_n W(T_3\cosh X_n+T_4\sinh X_n)
\end{equation}
where
\[X_n=\alpha n\sqrt A=2\frac{ v_0}{hc}n ,\quad
 W=\alpha\frac{ b B}{A\sqrt{A}}=\frac{w_0}{2c}\left(1-h^4\right)<0\]

\[
T_0=\frac{b}{A}+\frac{ a}{\sqrt{A}}W=\frac{1}{2}(1-h^2)(1+\frac{v_0}{c}(1+h^2))<0\]
\[ T_1=\frac{B}{A}-\frac{ a}{\sqrt{A}}W=\frac{1}{2}(1-h^2)(1-\frac{v_0}{c}(1+h^2))>0\]

\[
T_2=\frac{a}{\sqrt A}-W=h-\frac{v_0}{2hc}(1-h^4)>0,\quad  T_3=-\frac{ a^2}{b}=-\frac{2h^2}{1-h^2}>0\]

\[
  T_4=-\frac{Ba }{b\sqrt{A}}=-h\frac{1+h^2}{1-h^2}\approx h>0\]
The  term $\frac{aW}{\sqrt{A}}=-\frac{v_0}{c}\frac{(v_0^4-w_0^4)}{2w_0^4}=\frac{v_0}{2c}(1-h^4)$ marks the  sub dominant contributions; being proportional to $\frac{v_0}{c}$ it shows the importance of the velocity of propagation of the signal (quanta) and suggests that the next order of interaction  may require a proper relativistic approach. On the other hand  $\frac{v_0}{c}<<1$  does not assure  $|\frac{aW}{\sqrt{A}}|<1$, as long as there is no a priori constraint on the size of $h$.

\subsection*{Higher $n$-order contributions}
The coefficients $T_0$, $T_1$ and $T_2$ carry contributions from both the dominant and the sub dominant $n$-orders and they will receive contributions from any other lower $n$-order that may be added.  Considering the possibility of including the $N$ first $n$-orders, (\ref{xna2}) can be generalized to

\begin{equation}
\label{xnak}
x_n=T_0(N)+\sum_{k=0}^N \frac{(WX_n)^{k}}{2k!} \left(T_{2k+1}(N)\cosh X_n+T_{2k+2}(N)\sinh X_n\right).
\end{equation}
Every new $n$-order added adds contributions to every $T$-coefficient, including the $T_0$ one. It is a peculiar aspect of these $n$-order expansions that the relevance of lower $n$-order terms increases as $n$ grows
in contraposition to the usual approximation expansions in terms of small coupling constants where each order of contribution is fixed. 
This is unexpected as in general the gravitational effects decrease with the distance; these non-Newtonian effects increase. As the distance keeps increasing,  contributions from lower $n$-orders become more and more relevant. Probably there is a closed form, still to be found,  for each $T$-coefficient as well as for the entire equation of which the expansion (\ref{xnak}) would be just a partial sum. On the other hand there is no point on adding new contributions if they only become relevant with  distances that go beyond our observation horizon. 

\section{Discrete versus continuous }

Let us compare (\ref{xnh1}) and (\ref{xna2}). They represent two distinct perspectives, two different approaches to the dynamics of gravity. (\ref{xnh1}) stands for the non-relativistic weak-field limit of General Relativity, for the vision of an interaction field as continuous as a smooth surface while (\ref{xna2}) shows additional consequences of accumulated effects of successive discrete interactions. Let us now point to some striking  differences between them.

While (\ref{xnh1}) describes just one phase  evolution, (\ref{xna2}) describes a multi-phase one. The dynamics is very sensitive to the initial conditions.  $|W|<<1$ reduces (\ref{xna2}) to (\ref{xnh1}) as it turns the sub dominant $n$-order contributions negligible. On the other hand as $n$ (or equivalently the distance or the time) increases, the last two terms on the right-hand side of (\ref{xna2}), proportional to $WX_n$, and that we are calling  the sub dominant contributions, increases ($WX_n$ times)faster and may compete or even dominate, radically changing the dynamics.  Let 
\begin{equation}
\label{condN1}
 |W|X_{n}\approx 1\end{equation}  mark the point where these two contributions grow even; for $|W|X_{n}>1$  the non-Newtonian effects  increase faster than the Newtonian ones but for  $|W|X_{n}<<1$ 
(\ref{xna2})  reduces to 
\begin{equation}
\label{xna3}
x_n\approx T_0+T_1\cosh X_n+T_2\sinh X_n\qquad for \qquad |W|X_n<<1
\end{equation}
which is just (\ref{xnh1}) with different constants. The dynamics is
 strongly dependent on the initial conditions and, even after this  simplification, it still encompass a great variety of physical situations  of which we mention here some few. The same procedure applied to (\ref{xnh1}) leads \cite{else} now to a  distinct conservation law
\[q=p_n-\frac{\beta }{\alpha\sqrt{A}W}\ln\frac{(e^{X_n}-Z_{+})}{(e^{X_n}-Z_{-})}=p_0-\frac{\beta }{\alpha\sqrt{A}W}\ln\frac{(1-Z_{+})}{(1-Z_{-})}\]
with 
\begin{equation}
\label{Z}
Z_{\pm}(T_1+T_2)=-T_0\pm\sqrt{T_0^2+T_2^2-T_1^2}=-T_0\pm W\end{equation}
and leads to a dominant
 logarithmic contribution to the potential

\begin{equation}
\label{Un}
U_n\approx\frac{q\beta}{m\ W\sqrt{A}}\ln\left(\frac{x_n+\sqrt{x_n^2-2x_nT_0+W^2}-W}{x_n+\sqrt{x_n^2-2x_nT_0+W^2}+W}\right).
\end{equation}
For 
\begin{equation}
\label{Wm1}
|W|<<x_n
\end{equation}
both $q$ and $U_n$ reduce, in a first order approximation, to the Newtonian results, as expected.

\subsection*{Gravitation between elementary particles}\label{ssec9}

The  interesting result comes from the other extreme condition
\begin{equation}
\label{Wb1}
|W|>>x_n
\end{equation}
that reduces (\ref{Un}) to
\begin{equation}
\label{Un1}
U_n\approx-\frac{q\beta}{m\alpha W\sqrt{A}}\ln\left(1+\frac{2W}{x_n}\right)
\end{equation}
whose gradient produces an effective  ``acceleration"
\[a_n\approx -\frac{r_0}{m}\frac{dU_n}{dx_n}=-\frac{qGM}{ Ww_0}\frac{1}{ r_n}.\]
An $r_n$-dependence like that could explain the  flat rotation curves of galaxies. Besides, considering the conservative potential we can imagine, just to simplify, that a given star was carried to its actual state from an energetically equivalent initial position $r_0^{*}$ and an initial velocity  $v_0\approx0$, then the condition (\ref{Wb1}) reduces to

\[\frac{GM}{2c^2r_0^{*}}>>(\frac{r_n}{r_0^{*}})^2\] or\[a_{n,N}>>\frac{2c^2}{r_0^{*}}\] 
where $a_{n,N}=GM/r_n^2$ is the Newtonian value for the star acceleration at $r_n$ which must  be bigger than a reference value in a kind of twisted Milgron's  condition \cite{Milgrom}, in reference to an empirically found condition satisfied by most of flat rotation curves of galaxies.  
The nagging point is that plugging back the $N_M$ factor of (\ref{cNM})  turns the condition (\ref{Wb1}) into
\begin{equation}
\label{condN4}
(\frac{v_0h^3}{N_M c})^2\ge x_n^2
\end{equation}
which can be realized only with few-components objects like elementary particles
as they have small values of $N_M$ ( $N_M=1$), $x_n\approx1$,  excluding any macroscopic object, not to mention a star or galaxy.  It is saying that the gravitational interaction between two elementary particles could, in principle  (it depends on $h$, the initial conditions as the dynamics is very sensitive to them) radically differ from the one we know between macroscopic objects in usual conditions! Had we only a dream technology capable of measuring such discrepancies!

\section{Discrete interaction and cosmology}\label{ssec11}
The discrete interaction toy model considered in this paper can easily be associated to a Newtonian cosmological model of a spherically symmetrical distribution of dust of which $m$ is the mass of the point dust at $r_n$ and $M$ is the total mass of the dust spherically distributed inside the sphere of radius $r_n$. It is mostly a consensus that inflation, an early exponential expansion of the universe, is a necessary ingredient of big-bang cosmologies despite the many questions lingering on, waiting for consistent answers. Questions like, how did inflation started? powered by what? and      how did it smoothly ended into the present known state of the universe? Similar questions can also be made with respect to the observed accelerated expansion of the universe, that seems to require an evolution system with a repulsive gravitation phase.
Discrete interaction, we believe,  have something  to add  on these issues.

\subsection*{Exponential expansion  phases }\label{ssec10}

In the dynamics described by (\ref{xna2}) there are two main exponential expansion phases, an early and a late one. Let us start re-writing it as
\[x_n=T_0+\left(T_1+T_2+X_n W\frac{T_3+T_4}{2}\right)\frac{e^{X_n}}{2}+\]
\begin{equation}
\label{xna3a}
+\left(T_1-T_2+X_n W\frac{T_3-T_4}{2}\right)\frac{e^{-X_n}}{2}
\end{equation}
from which we can see that it describes an exponential expansion as long as
\begin{equation}
\label{xna4}
T_1+T_2+X_n W\frac{T_3+T_4}{2}>> T_1-T_2+X_n W\frac{T_3-T_4}{2}
\end{equation}
or

\begin{equation}
\label{xna4a}
X_{n}<<\frac{2T_2}{|W|T_4}
\end{equation}
and also when
\[X_{n}>>\frac{2(T_1+T_2)}{|W|(T_3+T_4)}\approx\frac{2T_1}{|W|T_4}>>\frac{2T_2}{|W|T_4}\]

as $T_1>>T_2$ and $T_4>>T_3.$ 
It is natural that if we want to use this discrete interaction model for pursuing a more comprehensive understanding of gravity  we must consider the whole (\ref{xna2}) and not just (\ref{xna3}). To repeat with (\ref{xna2}) the same treatment given to (\ref{xnh1}) and to (\ref{xna3}) is not as easy but we can do, for now, with numerical computation for a glance on its dynamics. See the figures below where we have used the command 

\[Manipulate[
 Plot[f(x,h), \{x, x_{min}, x_{max}\}], \{h, 
  h_{min}, h_{max}\}]\] of the software Mathematica 8.0 which 
is very useful  as it allows the  observation of the dynamics dependence on the initial conditions through the explicit variation of $h$. For the Figure 1, showing the repulsive, attractive and again repulsive sequence of phases,  we have plotted the "acceleration'' $v_n dv_n/dx_n $ against $X_n$  for (\ref{xna2}) with $A<0$ and with the initial conditions partially fixed  by taking $v_0=1000m/s$ and leaving $r_0$ free, hidden in the parameter $h$ which was allowed to vary in the range from $10^{-3}$ to $10^5$ and $x$ (that stands here for $X_n$) from 0 to $2\pi$. Increasing the range of $x$ would mean including distances beyond, perhaps, our observation horizon; for consistency this would require  including  more $n$-order contributions. A smaller $h$ implies, for a fixed initial velocity, a larger escape velocity which means, for a fixed $M$, a smaller initial position, while a larger $h$ implies $v_{E,0}$ closer to $v_0$.
For comparison, using (\ref{xnh1}) instead of (\ref{xna2}), for the same initial conditions, produces Figure 2 and an always negative Newtonian "acceleration''

\[v_n dv_n/dx_n = -4.5 \times10^6 (1 + 1/h^2)\left(\cos0.5X_n+h\sin0.5X_n\right)^{-4}\approx \]
\[-4.5 \times10^6 x_n^{-2}<0
\]
in its unique attractive phase.
\begin{figure}[tbp] 
  \centering
\includegraphics[bb=0 0 288 173,width=4.76in,height=4.66in,keepaspectratio]{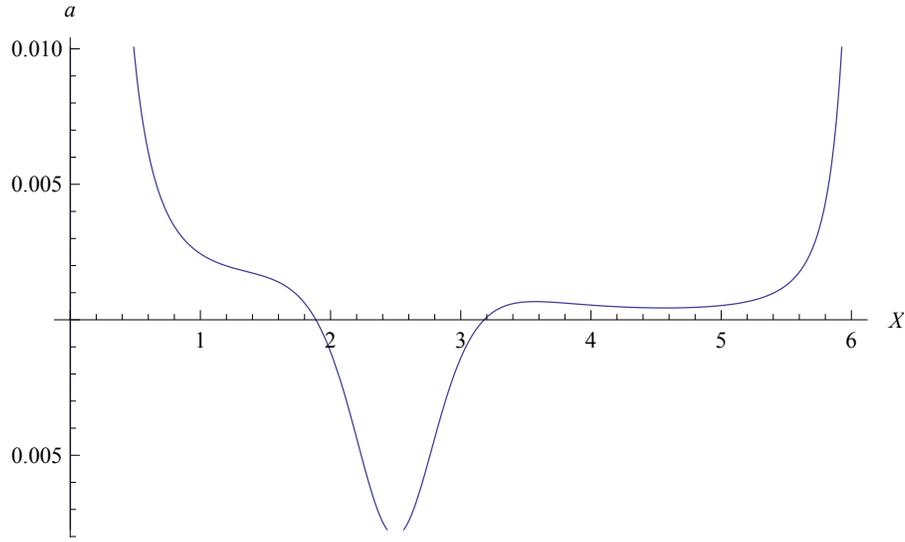}
  \caption{3 phases  from an infinite periodical series of repulsion, attraction and repulsion.}
  \label{fig:m111}
\end{figure}

\begin{figure}[tbp] 
  \centering
\includegraphics[bb=0 0 242 146,width=4.94in,height=4.77in,keepaspectratio]{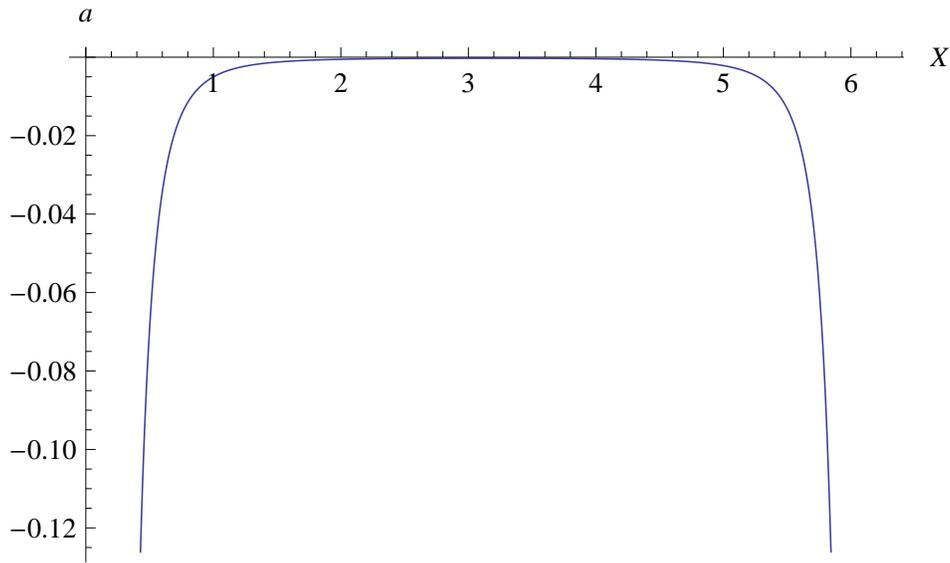}
  \caption{The unique Newtonian phase: $a_n<0.$}
  \label{fig:aNewt}
\end{figure}

We should observe that in both figures the scales of $X_n$ in the horizontal axis are not linear with the scale of distance $x_n$; they are related by (\ref{xnh1}) and (\ref{xna2}), respectively.
In the notation of  Mathematica , with  the chosen initial condition, we write (\ref{xna2}) as
\[x_n=((1 + h^2)/2 + 
   10^{-5} (1 - h^4)/2) +\]\[ ((1 - h^2)/2 - 10^{-5} (1 - h^4)/2 - 
    h 10^{-5} x (1 - h^2)/2) \cos[x] +\]\[ (h - 10^{-5} (1 - h^4)/2 + x 10^{-5} (1 - h^2)^2 /4) \sin[x].\]
Changing $h$ changes the overall aspect of the plot  but not the existence of changes of sign in the ``acceleration" as it is naturally dependent on the initial conditions. The choice of the snapshot taken at $h=360$  just reflects the author's esthetic bias without any further deep physical concern. 
\section{Conclusions}
The objective of  this paper  of pointing to  the relevance of discrete interaction models as laboratory of ideas in the search for a better comprehension of field interaction and of its  discrete to continuous or quantum to  classical transitions has been accomplished. Although being   based on a very simple   classical model of  discrete interactions for gravity and being mirrored on  the  non-relativistic weak field limit of General Relativity which is just an approximation to a correct formulation of the problem it makes contact with  actual major issues in the physics of gravity. Most remarkable, of course, is the natural appearing of repulsive gravity in a simple quasi-Newtonian context. The next question is how far  can it go.

\end{document}